\pdfoutput=1 
\documentclass[9pt, sigconf, natbib=false, screen, nonacm]{acmart}

\copyrightyear{2023} 
\acmYear{2023} 
\setcopyright{acmcopyright}
\acmConference[ASPDAC '23]{28th Asia and South Pacific Design Automation Conference}{January 16--19, 2023}{Tokyo, Japan}
\acmBooktitle{28th Asia and South Pacific Design Automation Conference (ASPDAC '23), January 16--19, 2023, Tokyo, Japan}
\acmPrice{15.00}
\acmDOI{10.1145/3566097.3567929}
\acmISBN{978-1-4503-9783-4/23/01}

\usepackage{tabularx}
\usepackage[frame,line,arrow,matrix,tips]{xy}	
\usepackage{color, colortbl}

\usepackage[T1]{fontenc}
\usepackage[utf8]{inputenc}
 
\usepackage{csquotes}
\usepackage{nicefrac}
\usepackage[textsize=tiny]{todonotes} 
\usepackage{booktabs}
\usepackage{tikz}
\usepackage{graphicx}
\usetikzlibrary{shapes.geometric, positioning, quantikz2, arrows, trees, fit}
\usepackage{yquant}
\usepackage{csvsimple}
\useyquantlanguage{groups}
\usepackage{nicematrix}
\usepackage{threeparttable}

\usepackage{xparse}
\ExplSyntaxOn
\DeclareExpandableDocumentCommand{\booltest}{mmm}
 {
  \bool_if:nTF { #1 } { #2 } { #3 }
 }
\DeclareExpandableDocumentCommand{\fpcompare}{m}
{
\fp_compare_p:n { #1 }
}
\DeclareExpandableDocumentCommand{\fptest}{m}
{
\fp_compare:nTF { #1 }
}
\ExplSyntaxOff

\makeatletter
\DeclareRobustCommand\rvdots{%
\vbox{%
\baselineskip4\p@\lineskiplimit\z@%
\kern-\p@%
\hbox{.}\hbox{.}\hbox{.}%
}%
}
\makeatother

\usepackage{subcaption}

\newtheorem{example}{Example}

\usepackage[style=ieee, maxnames=5, minnames=1, date=year, doi=false,isbn=false,backend=biber,dashed=false]{biblatex}
\AtEveryBibitem{
  \clearname{editor}%
  \clearfield{series}%
  \clearfield{isbn}%
  \clearfield{issn}%
    \clearfield{volume}%
  \clearfield{number}%
  \clearfield{pages}%
}

\usepackage[detect-all=true]{siunitx}
\usepackage{etoolbox}
\robustify\bfseries
\begin{document}

\title[A SAT Encoding for Optimal Clifford Circuit Synthesis]{A SAT Encoding for Optimal Clifford Circuit Synthesis}
\author{Sarah Schneider}
\affiliation{%
   \institution{Institute for Integrated Circuits,\\ Johannes Kepler University Linz, Austria}
   \city{}
   \country{}
}
\email{}

\author{Lukas Burgholzer}
\authornote{Corresponding author.}
\affiliation{%
   \institution{Chair for Design Automation,\\ Technical University of Munich, Germany}
   \city{}
   \country{}
}
\email{lukas.burgholzer@tum.de}

\author{Robert Wille}
\affiliation{%
   \institution{Chair for Design Automation,\\ Technical University of Munich, Germany}
   \city{}
   \country{}
}
\email{robert.wille@tum.de}

\begin{abstract}
	Executing quantum algorithms on a quantum computer requires compilation to representations that conform to all restrictions imposed by the device.
	Due to devices' limited coherence times and gate fidelities, the compilation process has to be optimized as much as possible.
	To this end, an algorithm's description first has to be \emph{synthesized} using the device's gate library.
	In this paper, we consider the \emph{optimal} synthesis of \emph{Clifford} circuits---an important subclass of quantum circuits, with various applications.
	Such techniques are essential to establish lower bounds for (heuristic) synthesis methods and gauging their performance.
	Due to the huge search space, existing optimal techniques are practically limited to small qubit counts (around six qubits for typical instances).
    In this work, we propose an optimal synthesis method for Clifford circuits based on \emph{encoding} the task as a satisfiability (\emph{SAT}) problem and solving it using a SAT solver in conjunction with a binary search scheme.
    Experiments on random instances with up to 6 qubits demonstrate that state-of-the-art heuristics on average produce more than twice the number of gates necessary.
\end{abstract}

\maketitle

\noindent\textbf{Correction notice.}

This version corrects the originally published ASPDAC'23 article (DOI: 10.1145/3566097.3567929).
A post-publication review revealed issues in the earlier evaluation pipeline that led to non-reproducible scalability claims.
We re-ran a thorough evaluation 
with the open-source implementation available in \href{https://github.com/munich-quantum-toolkit/qmap}{MQT QMAP}; the SAT-based method remains valid, but the reported scalability is corrected from 27 qubits to 6 qubits.
All figures were regenerated and all results are reproducible from the scripts and data in MQT QMAP (\href{https://github.com/munich-quantum-toolkit/qmap/tree/6a0d8a2ff411a0e2c9604c71aff80ba633c0d660/eval/clifford_state_prep}{@6a0d8a2}).

\section{Introduction}\label{sec:introduction}

In quantum computing, algorithms are generally described via quantum circuits---potentially consisting of high-level operations such as the quantum Fourier transform~\cite{nielsenQuantumComputationQuantum2010} or Grover iterations~\cite{groverFastQuantumMechanical1996}.
Alternatively, functional descriptions such as a unitary matrix, a decision diagram, or a tensor network can be used to describe the functionality of a particular algorithm~\cite{willeBasisDesignTools2022}.
Quantum computing architectures provide a certain set of \emph{native} gates, i.e., quantum gates that can be directly performed on the quantum computer.
These native gate-sets typically consist of a few single-qubit gates, as well as a particular two-qubit gate such as the CNOT---forming a universal set of gates~\cite{nielsenQuantumComputationQuantum2010}.
Similar to how, in classical computing, high-level C code has to be compiled to machine-dependent assembly, in quantum computing, an algorithm's high-level functional description has to be \emph{synthesized} using the respective gate library \cite{barencoElementaryGatesQuantum1995, maslovAdvantagesUsingRelative2016,willeImprovingMappingReversible2013,degriendArchitectureawareSynthesisPhase2020}.

Due to the limited fidelity and coherence times of (today's) quantum computers, it is critical that the synthesis is conducted as efficiently as possible.
Efficiency in this regard can relate to a variety of metrics, such as minimizing
the \emph{number of gates} or the \emph{depth} of the resulting circuit---roughly corresponding to the number of instructions or the runtime of the circuit, respectively.
More precisely, the number of \emph{two-qubit} gates (such as CNOT gates) has been identified as the main source for errors in currently available quantum systems and, thus, has been predominantly used as a cost metric for synthesizing circuits in the past.
Hence, we also focus on minimizing the two-qubit gate-count in the following.

In this work, we consider the question:
Given a quantum circuit, what is the most efficient realization of the respective functionality?
We particularly focus on the synthesis of \emph{Clifford} circuits, which form a central class of quantum circuits, important for error correcting codes \cite{gottesmanStabilizerCodesQuantum1997,devittQuantumErrorCorrection2013} and several important quantum phenomena, such as superposition, entanglement, superdense coding, as well as teleportation~\cite{nielsenQuantumComputationQuantum2010}.
Studying synthesis approaches that guarantee optimal solutions is important for establishing lower bounds on the achievable efficiency and for evaluating how far existing and future synthesis methods stray from this optimum. 
Even though Clifford circuits form a finite subgroup of all quantum circuits---one that is not even universal for quantum computing---the search space for the synthesis problem grows exponentially with respect to the number of considered qubits.

In this work, we propose to encode the optimal synthesis task as a satisfiability (\emph{SAT},~\cite{biereHandbookSatisfiability2009}) problem and solve it using a SAT solver in conjunction with a binary search scheme.
Besides their widespread use in the classical domain, these solvers have proven quite powerful in efficiently navigating the huge search spaces encountered in various problems related to quantum computing~\cite{willeMappingQuantumCircuits2019,tanOptimalLayoutSynthesis2020,berentSATEncodingQuantum2022,yamashitaFastEquivalencecheckingQuantum2010,matsuoReducingOverheadMapping2019,meuliSATbasedCNOTQuantum2018,burgholzerLimitingSearchSpace2022,willeATPGReversibleCircuits2011}.
Experimental results produced with the resulting tool---publicly available under the MIT license as part of the \emph{Munich Quantum Toolkit}~(MQT;~\cite{mqt}) in the QMAP tool~\cite{willeMQTQMAPEfficient2023} (\url{https://github.com/munich-quantum-toolkit/qmap})---demonstrate that the \mbox{state-of-the-art} heuristic for Clifford circuit synthesis (proposed in~\cite{bravyiCliffordCircuitOptimization2021} and available in IBM's Qiskit~\cite{qiskit2024}) produces on average more than twice the number of two-qubit gates than necessary on random instances with up to 6 qubits. 

The remainder of this work is structured as follows: 
\autoref{sec:background} gives a brief overview of quantum computing and Clifford circuits.
 \autoref{sec:synthesis} reviews the synthesis problem for quantum circuits and related work.
Then, \autoref{sec:cliffencoding} describes the proposed encoding.
Based on that, \autoref{sec:speedup} describes 
how to efficiently determine
optimal solutions.
Then, \autoref{sec:results} summarizes the experimental results, before \autoref{sec:conclusions} concludes the paper.

\section{Background}\label{sec:background}
In this section, we give a brief introduction to quantum computing and the main concepts needed throughout this work.
While the individual descriptions are kept brief, we refer to~\cite{nielsenQuantumComputationQuantum2010,aaronsonImprovedSimulationStabilizer2004,gottesmanStabilizerCodesQuantum1997} for a more in-depth introduction.

\subsection{Quantum Computing}\label{sec:quantum}
In classical computing, the basic unit of information is called a \emph{bit} and can assume either one of the two states $0$ and $1$.
In quantum computing, the basic unit of information is called a quantum bit (\emph{qubit}) and cannot only assume either one of two basis states, $\ket{0}$ and~$\ket{1}$, but also any \mbox{complex-valued} linear combination (\emph{superposition}) thereof, i.e., the state~$\ket{\psi}$ of a qubit is defined by $\ket{\psi} = \alpha_0 \ket{0} + \alpha_1 \ket{1}$, with $\alpha_0, \alpha_1 \in \mathbb{C}$ and $|\alpha_0|^2 + |\alpha_1|^2 = 1$.
The $\alpha_i$ are called amplitudes and are frequently written in the form of a state vector $\ket{\psi} \equiv \begin{bmatrix} \alpha_0 & \alpha_1 \end{bmatrix}^\top$.
Measuring (i.e., observing) the qubit causes its state to collapse to one of the basis states $\ket{i}$---each with probability $|\alpha_{i}|^2$.

The basis for an $n$-qubit state is formed by the tensor product of single-qubit states, i.e., $ \ket{i_{n-1}} \otimes \cdots \otimes \ket{i_0} \equiv \ket{i_{n-1} \dots i_0}$, with $i_j \in \{0,1\}$ for $j$ from $0$ to $n-1$.
Consequently, any $n$-qubit state can be described by $\ket{\psi} \equiv \sum_{i=0}^{2^{n}-1} \alpha_i\ket{i}$ with $\alpha_i \in \mathbb{C}$ and $\sum_{i=0}^{2^{n}-1} |\alpha_i|^2 = 1$.
One fundamental difference between classical and quantum states is \emph{entanglement}.
A quantum state is said to be entangled if the state of its qubits cannot be considered separately, but has to be considered as a whole.

\begin{example}\label{ex:bellstates}
	The four Bell states 
	\[\ket{\Phi^{\pm }} = 1/{\sqrt{2}} (\ket{00} \pm \ket{11}) \mbox{ and }\ket{\Psi^{\pm }} = 1/{\sqrt{2}} (\ket{01} \pm \ket{10})
	\]
	 are some of the most well-known examples of entangled states.
	Measuring, e.g., the first qubit of a system in the $\ket{\Psi^+}$ state results in either $\ket{01}$ or $\ket{10}$---each with probability $|1/{\sqrt{2}}|^2 = 0.5$.
	The state of the second qubit, even if not measured directly, depends on the state of the first one after the measurement.
\end{example}

In contrast to classical computing, any operation manipulating the state of a quantum system (a \emph{quantum gate}) is inherently reversible.
To this end, the functionality of a quantum gate acting on $k$ qubits is described by a \mbox{$2^k\times 2^k$-dimensional} unitary matrix~$U$.
Applying a quantum gate~$g$ to the state~$\ket{\psi}$ of a system corresponds to appropriately multiplying the corresponding unitary matrix~$U$ with the current state vector---resulting in a new state~$\ket{\psi'}$.

\begin{example}
	Some of the most fundamental single-qubit gates are the three Paulis, whose matrices are given by
	\[
		X = \begin{bmatrix}
			0 & 1 \\
			1 & 0
		\end{bmatrix},\qquad
		Y = \begin{bmatrix}
			0 & -i \\
			i & 0
		\end{bmatrix},\qquad
		Z = \begin{bmatrix}
			1 & 0  \\
			0 & -1
		\end{bmatrix}.
	\]
	In addition, the following three gates will be heavily used in the remainder of this work:
	\[
		H = \frac{1}{\sqrt{2}}\begin{bmatrix}
			1 & 1  \\
			1 & -1
		\end{bmatrix},\,
		S = \begin{bmatrix}
			1 & 0 \\
			0 & i
		\end{bmatrix},\mbox{ and }
		\mathit{CNOT} = \begin{bsmallmatrix}
			1 & 0 & 0 & 0 \\
			0 & 1 & 0 & 0 \\
			0 & 0 & 0 & 1 \\
			0 & 0 & 1 & 0
		\end{bsmallmatrix}.
	\]
\end{example}

\thispagestyle{empty}
A quantum circuit, like its classical counterpart, is then made up of a sequence of quantum gates.
This is conveniently described as a (quantum) circuit diagram, where circuit lines represent individual qubits and labeled boxes placed on them indicate the gates that are applied in sequence.
Executing a quantum circuit on some initial state $\ket{\psi}$ (typically assumed to be $\ket{0\dots 0}$) corresponds to successively applying the individual gates to the state. 

\begin{example}
	Consider again the four Bell states from \autoref{ex:bellstates}. Then, \autoref{fig:bellcircuit} shows the quantum circuits to generate these states from the all-zero state $\ket{00}$.
	First, $X$ gates are applied in order to prepare the various two-qubit basis states.
	Then, a Hadamard (H) gate is applied to the top qubit, followed by a CNOT controlled on the top qubit (indicated by $\bullet$) and targeted at the bottom qubit (indicated by $\oplus$)---eventually resulting in the Bell states shown earlier.
\end{example}

\begin{figure}[t]
	\centering
	\resizebox{\linewidth}{!}{
	\begin{tikzpicture}
		\node (a) {
		\begin{yquantgroup}
			\registers{
				qubit {} q[2];
			}
			\circuit{
				init {$\ket{0}$} q[0];
				init {$\ket{0}$} q[1];
				barrier (q);
				h q[0];
				cnot q[1] | q[0];
				output {$\ket{\Phi^+}$} (q[0-1]);}

			\circuit{
				init {$\ket{0}$} q[0];
				init {$\ket{0}$} q[1];
				x q[1];
				barrier (q);
				h q[0];
				cnot q[1] | q[0];
				output {$\ket{\Psi^+}$} (q[0-1]);}

		\end{yquantgroup}};
	
	\node[below=14mm of a.west, anchor=west] (b) {
		\begin{yquantgroup}
			\registers{
				qubit {} q[2];
			}
			\circuit{
				init {$\ket{0}$} q[0];
				init {$\ket{0}$} q[1];
				x q[0];
				barrier (q);
				h q[0];
				cnot q[1] | q[0];
				output {$\ket{\Phi^-}$} (q[0-1]);}

			\circuit{
				init {$\ket{0}$} q[0];
				init {$\ket{0}$} q[1];
				x q;
				barrier (q);
				h q[0];
				cnot q[1] | q[0];
				output {$\ket{\Psi^-}$} (q[0-1]);}

		\end{yquantgroup}};
	\end{tikzpicture}}\vspace*{-1mm}
	\caption{Quantum circuit generating the four Bell states}\label{fig:bellcircuit}\vspace*{-2mm}
\end{figure}

\subsection{Clifford Circuits\\ and the Stabilizer Formalism}\label{sec:cliffordgroup}
\emph{Clifford} circuits, i.e.,  quantum circuits generated from the set of gates $\lbrace H, S, \mathit{CNOT} \rbrace$, form one of the most important classes of quantum circuits \cite{nielsenQuantumComputationQuantum2010}. This is due to several factors:
\begin{itemize}
	\item According to the Gottesman-Knill theorem \cite{gottesmanHeisenbergRepresentationQuantum1998}, they can be simulated in polynomial time and space on classical computers using the \emph{stabilizer} formalism.
	\item They can be used to describe several quantum phenomena such as superposition, entanglement, superdense coding, and teleportation \cite{nielsenQuantumComputationQuantum2010}.
	\item Many error correcting codes rely on them \cite{gottesmanStabilizerCodesQuantum1997,devittQuantumErrorCorrection2013}.
\end{itemize}

The idea of the stabilizer formalism is to represent a quantum state not by a complex-valued vector of amplitudes, but by a set of operators that uniquely identify the state. A unitary operator $U$ is said to be a \emph{stabilizer} of a quantum state $\ket{\psi}$ if $U\ket{\psi} = \ket{\psi}$, i.e., $\ket{\psi}$ is an eigenvector of $U$ with eigenvalue 1.
\begin{example} 
	It is easy to see that the zero state $\ket{0}$ is stabilized by the Pauli $Z$ operator since $Z\ket{0} = \ket{0}$.
	The plus state, i.e., the state~$\ket{+}$ defined as $1/\sqrt{2} (\ket{0} + \ket{1})$, is stabilized by the Pauli $X$ operator since $X\ket{+}=\ket{+}$.
\end{example}
Quantum states that can be obtained from the all-zero basis state $\ket{0\dots0}$ by applying Clifford operations are called \emph{stabilizer states}.
The name originates from the fact that such a state is uniquely and efficiently described by the set of operators that generate the group of its stabilizers.
Specifically, any $n$-qubit stabilizer state can be described by a set of $n$ Pauli strings $\pm P_{i,0}P_{i,1}P_{i,2} \dots P_{i,n-1}$, with $P_{i,j} \in \lbrace I, X, Y, Z\rbrace$ and $i, j \in 0,\dots,n-1$.
Hence, two bits per qubit are needed to identify the Pauli operator, as well as one additional bit for the phase, which leads to a total of $n(2n+1)$ bits needed to uniquely describe a particular stabilizer state.

\begin{example}\label{ex:stabilizer}
	The stabilizer representation of a quantum state is conveniently described by a \emph{tableau}
	\[
				\begin{bNiceArray}{ccc|ccc|c}[]
					x_{0,0}   & \Cdots & x_{0,n-1}   & z_{0,0}    & \Cdots & z_{0,n-1}   & r_0    \\
					\Vdots    & \Ddots &  \Vdots         & \Vdots     & \Ddots &    \Vdots         & \Vdots \\
					x_{n-1,0} & \Cdots & x_{n-1,n-1} & z_{n-1,0} & \Cdots & z_{n-1,n-1} & r_{n-1}  \\
				\end{bNiceArray},
	\]
	or (in a more compact fashion) by
	\[
		\begin{bNiceArray}{ccc|ccc|c}[]
			\mathbf{x}_{0}   & \Cdots & \mathbf{x}_{n-1}   & \mathbf{z}_{0}    & \Cdots & \mathbf{z}_{n-1}   & \mathbf{r}    \\
		\end{bNiceArray}.
	\]
	Here, the binary variables $x_{ij}$ and $z_{ij}$ specify whether the Pauli term $P_{i,j}$ is $X$ or $Z$, respectively.
	Since $Y = iXZ$, setting $x_{ij} = z_{ij} = 1$ corresponds to $P_{i,j}=Y$.
	Finally, $r_i$ describes whether the generator has a negative phase.
\end{example}

The Gottesman-Knill theorem states that the generators of a stabilizer state can be updated in polynomial time after the application of a Clifford operation.
To this end, the following  update rules for the stabilizer tableau apply:
\begin{itemize}
	\item \emph{H} on qubit $j$: $\mathbf{x}_{j} \leftrightarrow \mathbf{z}_{j}$ and $\mathbf{r} \mathrel{\oplus}= \mathbf{x}_{j}\mathbf{z}_{j}$,
	\item \emph{S} on qubit $j$: $\mathbf{z}_{j} \mathrel{\oplus}= \mathbf{x}_{j}$ and $\mathbf{r} \mathrel{\oplus}= \mathbf{x}_{j}\mathbf{z}_{j}$, and
	\item \emph{CNOT} with control qubit $c$ and target qubit $t$, i.e., \\$\mathbf{x}_{t} \mathrel{\oplus}= \mathbf{x}_{c}$, $\mathbf{z}_{c} \mathrel{\oplus}= \mathbf{z}_{t}$, and $\mathbf{r} \mathrel{\oplus}= \mathbf{x}_{c}\mathbf{z}_{t}(\mathbf{x}_{t} \oplus \mathbf{z}_{c} \oplus 1)$, with $\oplus$ denoting the $XOR$ operation.
\end{itemize}
Using the stabilizer tableau in conjunction with these update rules allows for efficient simulation of Clifford circuits.

\thispagestyle{empty}
\begin{example}\label{ex:cliffordgroup}
	Consider the simulation of the circuit shown in \autoref{fig:circuit}.
	It starts off in the all-zero state $\ket{00}$, which has the stabilizer tableau representation
	$$
		\begin{bNiceArray}{cc|cc|c}
			0 & 0 & 1 & 0 & 0 \\
			0 & 0 & 0 & 1 & 0 \\
		\end{bNiceArray}\mbox{, with generators } \{ZI, IZ\} \mathrel{\hat{=}} \ket{00}.
	$$
	Applying the $H$ gate to qubit $0$ yields
	$$
		\begin{bNiceArray}{cc|cc|c}
			1 & 0 & 0 & 0 & 0 \\
			0 & 0 & 0 & 1 & 0 \\
		\end{bNiceArray}\mbox{, with generators } \{XI, IZ\} \mathrel{\hat{=}} \ket{0+}.
	$$
	Applying the $CNOT$ gate with control $0$ and target $1$ yields
	$$
		\begin{bNiceArray}{cc|cc|c}
			1 & 1 & 0 & 0 & 0 \\
			0 & 0 & 1 & 1 & 0 \\
		\end{bNiceArray}\mbox{, with generators } \{XX, ZZ\} \mathrel{\hat{=}} \ket{\Phi^+}.
	$$
	Eventually, applying both $H$ gates yields
	$$
		\begin{bNiceArray}{cc|cc|c}
			0 & 0 & 1 & 1 & 0 \\
			1 & 1 & 0 & 0 & 0 \\
		\end{bNiceArray}\mbox{, with generators } \{ZZ, XX\} \mathrel{\hat{=}} \ket{\Phi^+}.
	$$
	Hence, this circuit prepares the Bell state $\ket{\Phi^+}$ whose stabilizers are generated by $XX$ and $ZZ$.
\end{example} 

\begin{figure}[t]
	\centering
	\begin{subfigure}[t]{0.33\linewidth}
		\centering
		\begin{tikzpicture}
			\begin{yquant}
				qubit {$\ket{0}$} q[2];
				h q[1];
				cnot q[0] | q[1];
				h q[0];
				h q[1];
			\end{yquant}
		\end{tikzpicture}
		\caption{Using $H$ and $CNOT$}\label{fig:circuit}
	\end{subfigure}
	\hfill
	\begin{subfigure}[t]{0.32\linewidth}
		\centering
		\begin{tikzpicture}
			\begin{yquant}
				qubit {$\ket{0}$} q[2];
				h q[0];
				h q[1];
				[value=]
				phase q[1] | q[0];
				h q[1];
			\end{yquant}
		\end{tikzpicture}
		\caption{Using $H$ and $CZ$}\label{fig:circuit2}
	\end{subfigure}
	\hfill
	\begin{subfigure}[t]{0.32\linewidth}
		\centering
		\begin{tikzpicture}
			\begin{yquant}
				qubit {$\ket{0}$} q[2];
				h q[0];
				cnot q[1] | q[0];
			\end{yquant}
		\end{tikzpicture}
		\caption{Optimal circuit}\label{fig:circuit3}
	\end{subfigure}
	\caption{Three circuits implementing the same tableau}\label{fig:circuits}
\end{figure}

\section{Synthesis of Clifford Circuits}\label{sec:synthesis}
In this section, we review what it means to \emph{synthesize} Clifford circuits, why it is important to do this efficiently, and explore how this task has been conducted in the past.

\subsection{The Synthesis Problem}\label{sec:synthesis_problem}

\emph{Synthesis} is the task of determining a quantum circuit that implements a given functionality under certain constraints.
This functionality can be described in a multitude of ways, e.g., a (high-level) quantum circuit, a unitary matrix, a decision diagram, a tensor network, or---in the case of Clifford circuits---a stabilizer tableau.
The constraints typically originate from certain restrictions of actual quantum computers---predominantly their limited gate-set. 

\begin{example}\label{ex:target_tableau}
	Consider the final tableau from \autoref{ex:cliffordgroup}, i.e.,
	$$
		\begin{bNiceArray}{cc|cc|c}
			0 & 0 & 1 & 1 & 0 \\
			1 & 1 & 0 & 0 & 0 \\
		\end{bNiceArray}.
	$$
	As shown before, this tableau describes the Bell state $\ket{\Phi^+}$ and one possible circuit implementing this functionality has already been shown in~\autoref{fig:circuit}.
	\autoref{fig:circuit2} shows an alternative realization of the circuit using a controlled-$Z$ gate instead of the CNOT gate\footnote{The controlled-$Z$ gate is a Clifford operation since it can be constructed from a CNOT by surrounding its target with Hadamard gates.}.
\end{example}

In general, there are infinitely many ways to realize a given functional description as a quantum circuit.
However, due to the noisy nature of near-term quantum computers and the limited coherence times of their qubits, it is important to determine \emph{efficient} realizations.
In this regard, \emph{efficiency} can be defined in terms of the \emph{\mbox{gate-count}} or the resulting circuit's \emph{depth}---more or less translating to the number of instructions and the runtime of the quantum circuit. 
Specifically, a circuit is more efficiently synthesized if it has a lower number of gates or a lower depth. 

\begin{example}
	Consider again the tableau from the previous example. Then, the most efficient circuit implementing the desired functionality only consists of two gates---a Hadamard and a single CNOT---as shown in \autoref{fig:circuit3}.
	While, in this case, the difference between these three realizations is quite small, it can grow tremendously large in more practical scenarios (as, e.g., demonstrated later in~\autoref{sec:results}).
\end{example}

Most existing solutions (which are reviewed next), strive to optimize the two-qubit gate-count of the resulting circuits, which is well motivated by the fact that two-qubit gates (such as CNOT gates) are the predominant source of errors in quantum circuits executed on today's devices.
Thus, in this work, we also consider the two-qubit gate-count as the main objective.

\begin{figure*}[!t]
	\centering
	\begin{subfigure}[b]{0.4\linewidth}
		\centering
		\resizebox{\linewidth}{!}{
		\begin{tikzpicture}
			\begin{yquant}[register/default name=]
				qubit a[2];
				nobit wave;
				qubit b[2];

				init {$G =$} (a, wave, b);
				discard wave;

				text {$n$} (wave);
				text {$\rvdots$} (wave);
				box {} (a, wave, b);
				box {} (a, wave, b);
				text {$\dots$} (a);
				text {$O(n^2/log(n))$} (wave);
				text {$\dots$} (b);
				box {} (a, wave, b);
				box {} (a, wave, b);
			\end{yquant}
		\end{tikzpicture}}
	\end{subfigure}%
	\hfill
	\begin{subfigure}[b]{0.55\linewidth}
		\centering
		\resizebox{0.95\linewidth}{!}{
		\begin{tikzpicture}
			\node (a) {
			\begin{yquantgroup}[register/default name=]
				\registers{
					qubit a[1];
				}
				\circuit{
					init {$1$ qubit: } (a);

					[fill=white, draw=white]
					barrier a;
				}
				\circuit{
					h a;
				}
				\circuit{
					box {$S$} a;
				}
			\end{yquantgroup}
			};

			\node[below= of a.west, anchor=west] (b) {
			\begin{yquantgroup}[register/default name=]
				\registers{
					qubit b[2];
				}
				\circuit{
					init {$2$ qubits:} (b);
					[fill=white, draw=white]
					barrier b;
				}
				\circuit{
					h b[0];
					[fill=white, draw=white]
					barrier b[1];
				}
				\circuit{
					h b[1];
					[fill=white, draw=white]
					barrier b[0];
				}
				\circuit{

					box {$S$} b[0];
					[fill=white, draw=white]
					barrier b[1];
				}
				\circuit{
					box {$S$} b[1];
					[fill=white, draw=white]
					barrier b[0];
				}
				\circuit{
					cnot b[0] | b[1];
				}
				\circuit{
					cnot b[1] | b[0];
				}
			\end{yquantgroup}
			};
		
			\node[below=15mm of b.west, anchor=west] {
			\begin{yquantgroup}[register/default name=]
				\registers{
					qubit c[3];
				}
				\circuit{
					init {$3$ qubits:} (c);
					[fill=white, draw=white]
					barrier c;
				}
				\circuit{
					h c[0];
					[fill=white, draw=white]
					barrier c[1];
					[fill=white, draw=white]
					barrier c[2];
				}
				\circuit{
					h c[1];
					[fill=white, draw=white]
					barrier c[0];
					[fill=white, draw=white]
					barrier c[2];
				}
				\circuit{
					h c[2];
					[fill=white, draw=white]
					barrier c[1];
					[fill=white, draw=white]
					barrier c[0];
				}
				\circuit{

					box {$S$} c[0];
					[fill=white, draw=white]
					barrier c[1];
					[fill=white, draw=white]
					barrier c[2];
				}
				\circuit{
					box {$S$} c[1];
					[fill=white, draw=white]
					barrier c[0];
					[fill=white, draw=white]
					barrier c[2];
				}
				\circuit{
					box {$S$} c[2];
					[fill=white, draw=white]
					barrier c[1];
					[fill=white, draw=white]
					barrier c[0];
				}
				\circuit{
					cnot c[0] | c[1];
					[fill=white, draw=white]
					barrier c[2];
				}
				\circuit{
					cnot c[1] | c[0];
					[fill=white, draw=white]
					barrier c[2];
				}
				\circuit{
					cnot c[0] | c[2];
				}
				\circuit{
					cnot c[2] | c[0];
				}
				\circuit{
					cnot c[1] | c[2];
				}
				\circuit{
					cnot c[2] | c[1];
				}
			\end{yquantgroup}
			};
		\end{tikzpicture}}
	\end{subfigure}\vspace{-2mm}
	\caption{Search space of gate choices for the synthesis problem}\label{fig:search_space}
\end{figure*}

\subsection{Related Work and Motivation}\label{sec:related}

Determining efficient realizations is a hard problem.
Even for a finite group such as the Clifford group, the search space grows rapidly.
In fact, the size of the $n$-qubit Clifford group grows as~$2^{\Theta(n^2)}$.
One of the key results for efficiently synthesizing Clifford circuits has been proposed by Aaronson and Gottesman~\cite{aaronsonImprovedSimulationStabilizer2004}.
They introduce a canonical form for Clifford circuits and demonstrate how this canonical representation can be generated from an arbitrary stabilizer tableau.
An important corollary of their results is that any $n$-qubit Clifford circuit can be realized using at most $\mathcal{O}(n^2 / \log n)$ gates---establishing an (asymptotically-optimal) \emph{upper} bound on the required resources.
Besides that, heuristics such as \cite{niemannEfficientSynthesisQuantum2014,bravyiCliffordCircuitOptimization2021} have been proposed.

In order to establish \emph{lower} bounds on the achievable efficiency and to judge how well existing synthesis techniques perform, it is important to study synthesis approaches that guarantee \emph{optimal} solutions.
Due to the immense search space, the number of qubits for which optimal Clifford circuits are known is shockingly small. 
In~\cite{kliuchnikovOptimizationCliffordCircuits2013}, optimal Clifford circuits with up to four inputs are reported while, to the best of our knowledge, the largest number of qubits for which optimal realizations have been demonstrated is \emph{six}. 
This has been shown by \emph{Bravyi et al.}~\cite{bravyi6qubitOptimalClifford2020}, where a pruned \mbox{breadth-first} search scheme is used to generate a database of classes of Clifford unitaries with known optimal canonical representatives---complemented by efficient means to search the created database\footnote{For reference, this database took $6$ months to generate and requires $\SI{2.1}{\tera\byte}$ of space.}.

In this work, we propose an alternative approach using satisfiability techniques (SAT,~\cite{biereHandbookSatisfiability2009}) to encode the synthesis problem (as described next in \autoref{sec:cliffencoding}) and solve it optimally using a SAT solver in conjunction with a binary search scheme (as described later in \autoref{sec:speedup}).
SAT techniques have seen great success in the design of classical circuits and are increasingly getting used to efficiently navigate the huge search spaces of various problems in quantum computing~\cite{willeMappingQuantumCircuits2019,tanOptimalLayoutSynthesis2020,berentSATEncodingQuantum2022,yamashitaFastEquivalencecheckingQuantum2010,matsuoReducingOverheadMapping2019,meuliSATbasedCNOTQuantum2018,burgholzerLimitingSearchSpace2022,willeATPGReversibleCircuits2011}.

\section{Encoding Clifford Circuits \\for Synthesis}\label{sec:cliffencoding}

In the following, we construct a generic satisfiability encoding for the functionality of a Clifford circuit, i.e, the gates and the corresponding tableau.
Furthermore, we show how the synthesis problem can be formulated based on this encoding and how the cost metric is specified.

\subsection{Encoding the Functionality\\ of a Clifford Circuit}\label{sec:circuit_encoding}

\thispagestyle{empty}
As reviewed in \autoref{sec:related}, any $n$-qubit Clifford circuit can be realized using $\mathcal{O}(n^2/\log n)$ gates\footnote{A precise upper limit $T$ is explicitly given by the canonical form of Aaronson and Gottesman~\cite{aaronsonImprovedSimulationStabilizer2004}.}.
Each of these gates can either be a single-qubit $H$ or $S$ gate on any qubit, or a two-qubit CNOT between any pairs of qubits---spanning the search space for the synthesis.
Thus, overall, at any point in time, there are
$$
	1 + 2n + 2\cdot n (n-1)/2  = 1 + n + n^2 = \mathcal{O}(n^2)
$$
possible choices for gates.
As a consequence, given a certain \mbox{time-step} limit $T\in\mathcal{O}(n^2/\log n)$,
a total of $T (1+n+n^2) \in \mathcal{O}(n^4 / \log n)$ Boolean variables suffice to describe all possible gates implementing a tableau, i.e., the structure of an arbitrary $n$-qubit Clifford circuit. 

\begin{example}\label{ex:search_space}
	\autoref{fig:search_space} illustrates the structure of the search space (i.e., the gate-variables) for the synthesis problem and shows all different choices of gates for up to three qubits:
	\begin{itemize}
		\item For $n=1$, there are just three options---doing nothing, apply a Hadamard gate, or applying an $S$ gate.
		\item For $n=2$, the single-qubit possibilities are complemented by two possible applications of a CNOT gate---resulting in a total of seven possible choices.
		\item For $n=3$, this number increases to thirteen.
	\end{itemize}
	In the resulting encoding, each possible placement of a gate at each \mbox{time-step} is represented by a corresponding gate variable.
\end{example}

Based on that, the functionality of a Clifford circuit for a given input state can be encoded by representing a stabilizer tableau for every considered \mbox{time-step}.
Since each of these tableaus requires to store $n(2n+1)$ bits, another set of $T(n(2n+1))\in \mathcal{O}(n^4 / \log n)$ Boolean variables are needed---one for each bit in a tableau at each \mbox{time-step}, as can be seen in \autoref{ex:stabilizer}.
 
In addition, functional constraints need to be added to ensure the consistency between the gate and the tableau variables. These have the form:
\begin{align*}
	 & \forall \mbox{ gate } g \mbox{ of the circuit } \hspace{3.5cm} {\color{gray}T\in\mathcal{O}(n^2 / \log n)}                           \\
	 & \forall \mbox{ choice } c \mbox{ for gate } g \hspace{4.6cm} {\color{gray}1 + n + n^2}                                      \\
	 & \forall \mbox{ row } r \mbox{ of the tableau for gate } g \hspace{4.1cm} {\color{gray}n}                            \\
	 & \resizebox{0.98\linewidth}{!}{$\mbox{update row } r \mbox{ of tableau for gate } g \mbox{ depending on gate choice } c,$}
\end{align*}
where each of these row updates corresponds to one of the tableau update rules reviewed in \autoref{sec:cliffordgroup}.
This results in a total of $T(1+n+n^2)n\in\mathcal{O}(n^5/\log n)$ constraints to ensure the consistency between the gate and the tableau variables.
Furthermore, cardinality constraints on the gate variables within a time-step ensure that only a single gate is executed per \mbox{time-step}.

\begin{figure*}
	\centering
	\resizebox{0.6\linewidth}{!}{
		\begin{tikzpicture}
			\begin{yquant}[register/default name=]
				qubit a[2];
				[name=wave, register/minimum height=5mm]
				nobit wave;
				qubit b[2];

				box {$\begin{bNiceArray}{c|c|c}[hvlines-except-borders]
								0 & I & 0
							\end{bNiceArray}$\\\\
						Fixed initial\\state $\ket{0\cdots 0}$} (a, wave, b);
				text {$n$} (wave);
				text {$\rvdots$} (wave);
				box {$\begin{bNiceArray}{c|c|c}[hvlines-except-borders]
								\ddots& \ddots & \ddots
							\end{bNiceArray}$\\\\
						$O(n^2)$ choices with\\$O(n)$ constraints} (a, wave, b);
				text {$\dots$} (a);
				text {$\times $ allowed\\ \mbox{gate-count}} (wave);
				text {$\dots$} (b);
				box {$\begin{bNiceArray}{c|c|c}[hvlines-except-borders]
								\ddots& \ddots & \ddots
							\end{bNiceArray}$\\\\
						$O(n^2)$ choices with\\$O(n)$ constraints} (a, wave, b);
				box {$\begin{bNiceArray}{c|c|c}[hvlines-except-borders]
								X & Z & r
							\end{bNiceArray}$\\\\
						Target tableau\\of $G$} (a, wave, b);
			\end{yquant}
		\end{tikzpicture}
	}
	\vspace{-4mm}
	\caption{Formulation of the synthesis problem based on the proposed encoding}\label{fig:encoding}
	\vspace*{-0.75em}
\end{figure*}

\begin{example}\label{ex:variable_constraints}
	Assume that $n=2$, as, e.g., for the circuits shown in \autoref{fig:circuits}.
	Then, an additional $n(2n+1) = 10$ Boolean variables are required per considered gate to encode the respective stabilizer tableau.
	As per \autoref{ex:search_space}, there are $7$ possible choices per gate (of which only one can be chosen at a time).
	This leads to a total of $7\cdot 2 = 14$ functional and one cardinality constraint per gate.
\end{example}

\subsection{Formulating the Synthesis Problem}

Based on the encoding proposed above, the synthesis problem can be formulated as illustrated in \autoref{fig:encoding}.
Given an $n$-qubit Clifford circuit, the synthesis starts off by using the polynomial algorithm proposed by Gottesman and Knill~\cite{gottesmanHeisenbergRepresentationQuantum1998} to construct what we refer to as the \emph{target tableau} in the following (shown on right-hand side of \autoref{fig:encoding}).
Alternatively, the target tableau can be explicitly given as input to the synthesis method.
Starting from the tableau for the all-zero state $\ket{0\dots 0}$ (shown on the left-hand side of \autoref{fig:encoding}), the goal is to determine an assignment of the encoding's variables (identifying a Clifford circuit) that realizes this target tableau. 

In order to efficiently guide the synthesis and to judge the quality of a solution, a cost metric (or objective function) 
is specified.
As discussed in \autoref{sec:synthesis_problem}, the main focus in this work is minimizing the number of two-qubit gates which are considered the predominant source of errors.
In the encoding proposed above, the number of gates in the resulting circuit corresponds to the respective \mbox{gate-choice} variables which are not assigned the identity.
Naturally, this can be limited further to only consider two-qubit gates.
The cost metric adds one cost constraint for each (two-qubit) \mbox{gate-choice} at any time-step, i.e., $Tn^2\in\mathcal{O}(n^4/\log n)$ constraints in total.

\begin{example}
	Considering the circuits shown in \autoref{fig:circuits} and, for the sake of simplicity, assuming equal costs for all gates, the following costs arise:
	The circuits in \autoref{fig:circuit} and \autoref{fig:circuit2} have a cost of four, while the circuit from \autoref{fig:circuit3} has a cost of two.
	Due to this, the last circuit is the best of these three with regard to the \mbox{gate-count} (in fact it is the optimal solution among all two-qubit Clifford circuits realizing the considered target tableau).
\end{example}

\section{The Resulting Optimal \\Synthesis Method}\label{sec:speedup}

As discussed in \autoref{sec:related}, optimal synthesis approaches are crucial for establishing lower bounds for the achievable efficiency and evaluating the performance of heuristic techniques.
The encoding proposed in the previous section is naturally suited for determining optimal solutions to the synthesis problem using SAT or SMT solvers.
Given a target tableau,
determining an optimal circuit realizing this tableau corresponds to constructing a SAT instance based on the proposed encoding and then minimizing the given cost function.

\subsection{Finding an Initial Time-Step Limit}\label{sec:initialvalues}

\thispagestyle{empty}
The runtime of the synthesis significantly depends on the number of \mbox{time-steps} considered, since
$\mathcal{O}(n^2)$ variables and $\mathcal{O}(n^3)$ constraints are generated for each \mbox{time-step}.
Thus, it is critical to find a suitably low initial guess for this number.
A straight-forward solution would be to take the minimum of the cost of the input circuit and the upper bound given by the canonical form of Aaronson and Gottesman~\cite{aaronsonImprovedSimulationStabilizer2004}.
While this is guaranteed to produce a satisfiable solution, i.e., the optimum is guaranteed to be in the search space of the SAT solver, it might result in an encoding that is excessively large.

Instead, the proposed encoding itself can be used to determine a more efficient initial guess by starting with a small initial limit (e.g., $T=n$) and letting the SAT solver search for \emph{any} solution within that time limit.
In general, the search for a solution terminates very quickly whenever no feasible solution exists given the prescribed limit.
Thus, the limit is gradually increased in a geometric fashion until a feasible solution is found.

\subsection{Performing the Optimization}\label{sec:optimization}
Once a suitable \mbox{time-step} limit $T$ has been determined---either by the method described above, from the original circuit itself, or the canonical form of Aaronson and Gottesman---there are two options for performing the actual optimization:

In standard SAT encodings, each constraint added to the problem instance is regarded a \emph{hard} constraint, i.e., it needs to be satisfied at all times in order for the obtained solution to be valid.
Using these standard capabilities of a SAT solver to tackle the underlying optimization problem, a binary search scheme can be used to determine the optimal solution by iteratively adapting the \mbox{time-step} limit~$T$. 
The value $T$ is adapted based on the halfway distance between an upper and a lower limit---in the first iteration, the upper limit is given by the canonical form of Aaronson and Gottesman, while the lower bound is 1. Upon achieving satisfiability with a given $T$, the new upper limit is set to be the current $T$. 
Should a $T$ not produce a satisfiable solution, the lower limit is set to be the current $T$. 
The process is repeated until a solution can be found for $T$ \mbox{time-steps} but not for $T-1$ \mbox{time-steps}---yielding the \emph{optimal} solution.

In contrast, MaxSAT formulations also accept \emph{soft} constraints which need not be satisfied necessarily.
The goal of corresponding solvers is to \emph{maximize} the number of satisfied soft constraints.
Soft constraints generally are additionally associated with a weight, which is used to determine the relative importance of the constraint. 

Overall, both procedures result in synthesis methods that allow to determine \emph{optimal} realizations of Clifford circuits---with a clear tradeoff between them.
While the MaxSAT approach automatically handles the optimization, the binary search scheme requires multiple executions of the SAT solver and iterative tuning of the time-step limit.
On the other hand, each individual call to the SAT solver is significantly less complex compared to the MaxSAT procedure.
As demonstrated in the following evaluations, the binary search scheme typically runs faster than the MaxSAT variant on the cases considered in this work.

\section{Experimental Results}
\label{sec:results}

The observations and resulting strategies proposed above can be used with any SAT-engine that allows for optimization and the inclusion of certain SMT theories, such as Z3~\cite{demouraZ3EfficientSMT2008} or \mbox{OptiMathSAT~\cite{sebastianiOptiMathSATToolOptimization2020}}.
Based on the encoding in \autoref{sec:cliffencoding}, the methods proposed in \autoref{sec:speedup} have been implemented in C++ (using the SMT-solver Z3~\cite{demouraZ3EfficientSMT2008}) and are publicly available under the MIT license as part of the \emph{Munich Quantum Toolkit} (MQT;~\cite{mqt}) in the QMAP tool~\cite{willeMQTQMAPEfficient2023} (\url{https://github.com/munich-quantum-toolkit/qmap}).
The respectively obtained results are summarized in the following.

All evaluation scripts and plotting code are available in the public QMAP repository. The full evaluation pipeline is scripted end-to-end and can be run from the repository's Clifford synthesis experiments; figures are generated directly from the exported evaluation data without manual curation.
We used MQT QMAP (v3.3.1), Qiskit (v2.1.2), as well as z3 (v4.15.3) on Python 3.12.9 and executed the experiments on an Apple M2 MacBook (macOS Sequoia, ARM64, 12 logical cores, 64 GiB unified memory).

For the benchmarks, random Clifford instances are generated with Qiskit and stored as QPY files.
For each number of qubits $n \in \{1,2,3,4,5,6\}$, we select various target sizes (i.e., numbers of gates) and sample 100 instances per target size.
Target sizes follow: for $n\le 4$, we use $\{1,n,2n,5n,10n\}$; for $n=5$, we use $\{1,5,10,15,20\}$; and for $n=6$, we use $\{1,6,9,12,15\}$.

We consider two baselines: the Aaronson--Gottesman (AG) synthesis~\cite{aaronsonImprovedSimulationStabilizer2004} and the greedy heuristic synthesis based on \cite{bravyiCliffordCircuitOptimization2021} and implemented in Qiskit (Greedy).
We compare these to four QMAP variants: two using the iterative binary search scheme (QMAP-Iter) and two using the MaxSAT formulation (QMAP-MaxSAT).
Each of these pairs differ in their initialization: one starts from a tableau (QMAP-Iter/Tableau and QMAP-MaxSAT/Tableau) using an initial timestep limit set to $\min(|\mathrm{AG}|,|\mathrm{Greedy}|)$, while the other starts from the better of AG/Greedy (QMAP-Iter/Circuit and QMAP-MaxSAT/Circuit).
All QMAP variants yield the same gate counts for a given instance; they only differ in runtime.

\begin{figure}[t]
    \centering
    \includegraphics[width=0.9\linewidth]{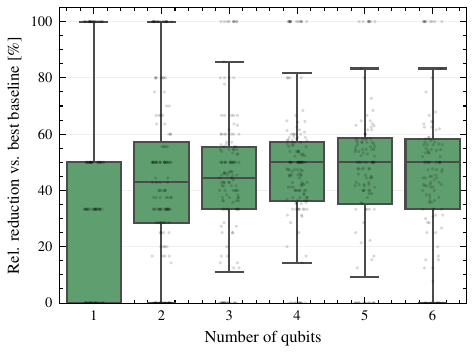}\vspace*{-4mm}
    \caption{Relative gate-count reduction of the proposed method versus the best baseline (min of AG/Greedy).}
    \label{fig:gate-improvement-boxplot}
\end{figure}

\begin{figure*}[t]
\centering
\includegraphics[width=\textwidth]{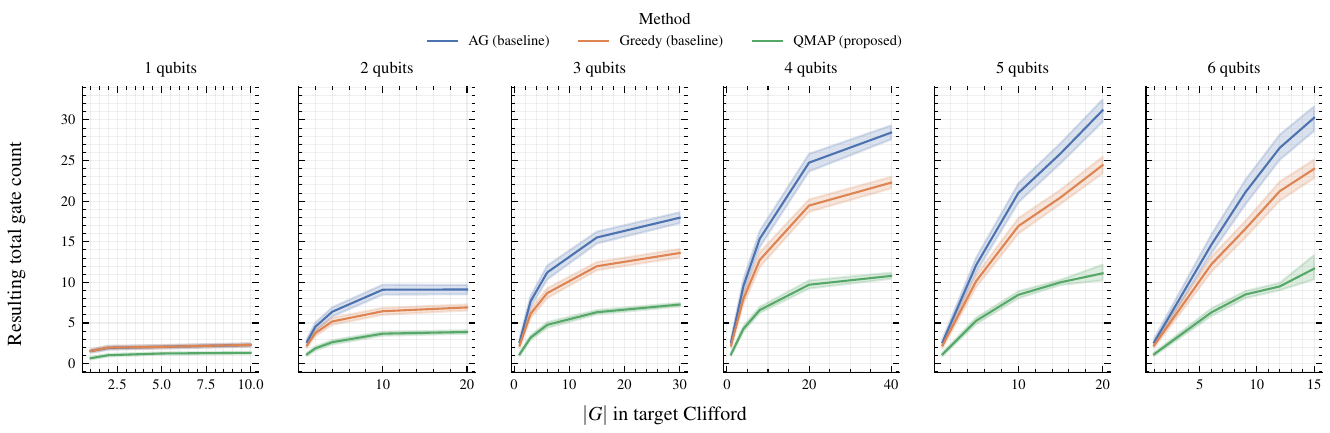}\vspace*{-5mm}
\caption{Gates vs. gates: AG and Greedy baselines compared to the aggregated QMAP result (means with 95\% confidence intervals), faceted by qubit count.
Results demonstrate that QMAP consistently improves over the baselines, with larger improvements for larger circuits.
}
\label{fig:gates-vs-gates}\vspace*{-2mm}
\end{figure*}

\begin{figure*}[t]
\centering
\includegraphics[width=\textwidth]{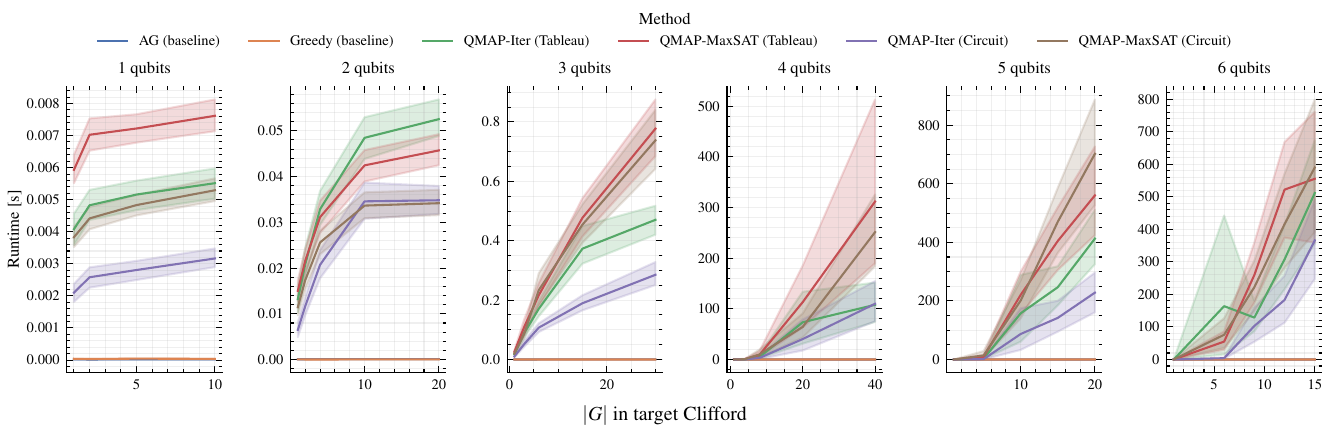}\vspace*{-5mm}
\caption{Runtime comparison versus target gate size, faceted by qubit count (only instances solved by all methods are shown).
Results demonstrate that QMAP-Iter is consistently faster than QMAP-MaxSAT.
The initialization from a circuit is faster than from a tableau.
}
\label{fig:runtime-vs-gates}\vspace*{-2mm}
\end{figure*}

For each combination of $(n, \text{target size}, \text{instance}, \text{method})$, we run the synthesis and record (i) the runtime in seconds and (ii) the total gate count.
All experiments are run with a hard per-run timeout of \SI{1800}{\second}.
The results are summarized in \autoref{fig:gates-vs-gates}, \autoref{fig:runtime-vs-gates}, and \autoref{fig:gate-improvement-boxplot}.

\paragraph{Completion rates and timeouts.} Each synthesis run used a hard timeout of \SI{1800}{\second}.
While both baselines (AG and Greedy) finished all runs within fractions of a second, the SAT-based methods did not solve all instances within the time limit—particularly for larger numbers of qubits and deeper circuits.
The iterative variants completed substantially more runs than their MaxSAT counterparts, and initialization from a circuit generally finished more runs than initialization from a Clifford tableau.
A detailed breakdown is provided in \autoref{tab:completion}.

\begin{table}[t]
    \centering
    \caption{Completion status (\# solved per 100 instances) across methods and settings. Baselines: AG (Aaronson--Gottesman) and Greedy (Qiskit). QMAP variants: Iter/MaxSAT with initialization from a tableau (Tab.) or a circuit (Circ.).}
    \label{tab:completion}
    \scriptsize
    \resizebox{\linewidth}{!}{
    \begin{tabular}{ccrrrrrr}
        \toprule
        Qubits & Gates & AG & Greedy & Iter (Tab.) & MaxSAT (Tab.) & Iter (Circ.) & MaxSAT (Circ.) \\
        \midrule
        1 & 1  & 100/100 & 100/100 & 100/100 & 100/100 & 100/100 & 100/100 \\
        1 & 2  & 100/100 & 100/100 & 100/100 & 100/100 & 100/100 & 100/100 \\
        1 & 5  & 100/100 & 100/100 & 100/100 & 100/100 & 100/100 & 100/100 \\
        1 & 10 & 100/100 & 100/100 & 100/100 & 100/100 & 100/100 & 100/100 \\
        2 & 1  & 100/100 & 100/100 & 100/100 & 100/100 & 100/100 & 100/100 \\
        2 & 2  & 100/100 & 100/100 & 100/100 & 100/100 & 100/100 & 100/100 \\
        2 & 4  & 100/100 & 100/100 & 100/100 & 100/100 & 100/100 & 100/100 \\
        2 & 10 & 100/100 & 100/100 & 100/100 & 100/100 & 100/100 & 100/100 \\
        2 & 20 & 100/100 & 100/100 & 100/100 & 100/100 & 100/100 & 100/100 \\
        3 & 1  & 100/100 & 100/100 & 100/100 & 100/100 & 100/100 & 100/100 \\
        3 & 3  & 100/100 & 100/100 & 100/100 & 100/100 & 100/100 & 100/100 \\
        3 & 6  & 100/100 & 100/100 & 100/100 & 100/100 & 100/100 & 100/100 \\
        3 & 15 & 100/100 & 100/100 & 100/100 & 100/100 & 100/100 & 100/100 \\
        3 & 30 & 100/100 & 100/100 & 100/100 & 100/100 & 100/100 & 100/100 \\
        4 & 1  & 100/100 & 100/100 & 100/100 & 100/100 & 100/100 & 100/100 \\
        4 & 4  & 100/100 & 100/100 & 100/100 & 100/100 & 100/100 & 100/100 \\
        4 & 8  & 100/100 & 100/100 & 100/100 & 100/100 & 100/100 & 100/100 \\
        4 & 20 & 100/100 & 100/100 & 99/100  & 98/100  & 94/100  & 95/100  \\
        4 & 40 & 100/100 & 100/100 & 97/100  & 96/100  & 97/100  & 98/100  \\
        5 & 1  & 100/100 & 100/100 & 100/100 & 100/100 & 100/100 & 100/100 \\
        5 & 5  & 100/100 & 100/100 & 100/100 & 100/100 & 100/100 & 100/100 \\
        5 & 10 & 100/100 & 100/100 & 100/100 & 88/100  & 98/100  & 86/100  \\
        5 & 15 & 100/100 & 100/100 & 85/100  & 66/100  & 85/100  & 61/100  \\
        5 & 20 & 100/100 & 100/100 & 56/100  & 38/100  & 58/100  & 36/100  \\
        6 & 1  & 100/100 & 100/100 & 100/100 & 100/100 & 100/100 & 100/100 \\
        6 & 6  & 100/100 & 100/100 & 99/100  & 97/100  & 97/100  & 97/100  \\
        6 & 9  & 100/100 & 100/100 & 88/100  & 78/100  & 95/100  & 74/100  \\
        6 & 12 & 100/100 & 100/100 & 60/100  & 49/100  & 75/100  & 41/100  \\
        6 & 15 & 100/100 & 100/100 & 39/100  & 20/100  & 55/100  & 22/100  \\
        \midrule
        TOTAL &  & 2900/2900 & 2900/2900 & 2723/2900 & 2630/2900 & 2754/2900 & 2610/2900 \\
        \bottomrule
    \end{tabular}}
\end{table}

\section{Conclusions}
\label{sec:conclusions}

In this work, we proposed a method for the optimal synthesis of Clifford circuits by encoding the task as a satisfiability problem and solving it via a binary search scheme.
The implementation is publicly available as part of the Munich Quantum Toolkit (MQT;~\cite{mqt}) in the QMAP tool~\cite{willeMQTQMAPEfficient2023} (\url{https://github.com/munich-quantum-toolkit/qmap}).
Overall, the resulting tool provides optimal solutions for small instances (1--6 qubits in our experiments) and, thus, establishes essential lower bounds for evaluating heuristic synthesis algorithms.
Future work includes exploring different cost metrics, extracting scalable heuristics from the proposed formulation, and extending beyond Clifford circuits.

\section*{Acknowledgments}
The authors would like to thank R. Kueng for his valuable advice and insightful discussions over the course of this project.

This work received funding from the European Research Council (ERC) under the European Union’s Horizon 2020 research and innovation program (grant agreement No. 101001318), was part of the Munich Quantum Valley, which is supported by the Bavarian state government with funds from the Hightech Agenda Bayern Plus, and has been supported by the BMWK on the basis of a decision by the German Bundestag through project QuaST, as well as by the BMK, BMDW, and the State of Upper Austria in the frame of the COMET program (managed by the FFG).

\printbibliography
\end{document}